\documentclass{ws-ijmpcs}
\newcommand{\be}{\begin{equation}}
\newcommand{\ee}{\end{equation}}
\newcommand{\bdm}{\begin{displaymath}}
\newcommand{\edm}{\end{displaymath}}
\def\dmf{\dot{\mathfrak{M}}}

\begin{document}

 \markboth{NAZAR IKHSANOV, NINA BESKROVNAYA and YURY LIKH}
 {MAGNETO-LEVITATION ACCRETION IN X-RAY PULSARS}

\catchline{}{}{}{}{}
%
%%%%%%%%%%%%%%%%%%%%%%%%%%%%%%%%%%%%%%%%%%%%%%%%%%%%%%%%%%%%%%%%%%%%

\title{EVIDENCE FOR MAGNETO-LEVITATION ACCRETION IN LONG-PERIOD X-RAY PULSARS}

\author{NAZAR IKHSANOV\footnote{Also the
Astrophys. Dept., Saint-Petersburg State University, Universitetskij pr.\,28, St.\,Petersburg, 198504 Russia},
NINA BESKROVNAYA and YURY LIKH}

\address{Pulkovo Observatory, Pulkovskoe Shosse 65--1,\\
 St. Petersburg, 196140 Russia
\\
ikhsanov@gao.spb.ru}

%\author{NINA BESKROVNAYA}

% \address{Pulkovo Observatory, Pulkovskoe Shosse 65--1,\\
% St. Petersburg, 196140 Russia
% \\
% beskrovnaya@yahoo.com}
%
%\author{Yury Likh}
%
%\address{Pulkovo Observatory, Pulkovskoe Shosse 65--1,\\
% St. Petersburg, 196140 Russia
% \\
% yury.likh@gmail.com}

\maketitle

\begin{history}
\received{30 Oct 2013}
%\revised{Day Month Year}
\end{history}

\begin{abstract}
Study of observed spin evolution of long-period X-ray pulsars challenges quasi-spherical
and Keplerian disk accretion scenarios. It suggests that the magnetospheric radius of
the neutron stars is substantially smaller than Alfv\'en radius and the spin-down torque
applied to the star from accreting material significantly exceeds the value predicted
by the theory. We show that these problems can be avoided if the fossil magnetic field of
the accretion flow itself is incorporated into the accretion model. The initially
spherical flow in this case decelerates by its own magnetic field and converts into a
non-Keplerian disk (magnetic slab) in which the material is confined by its intrinsic
magnetic field (``levitates'') and slowly moves towards the star on a diffusion
timescale. Parameters of pulsars expected within this magneto-levitation accretion
scenario are evaluated.

\end{abstract}

\ccode{PACS numbers: 97.60.Jd, 97.10.Gz, 97.10.Ld, 97.60.Gb}

  \section{Introduction}

It is widely believed that the geometry of accretion flow onto a neutron star in High-Mass X-ray Binaries (HMXBs) can be treated in either quasi-spherical or Keplerian disk approximations. The maximum possible spin-down rate which a neutron star could achieve within these  scenarios is, however, an order of magnitude smaller than the spin-down rate occasionally observed in the accretion-powered pulsars. This may indicate that  traditionally used accretion scenarios are oversimplified. We find that the above mentioned problem can be avoided if the magnetic field of the accreting material is incorporated into the accretion model. The star in this case is surrounded by and accreting from a magnetized non-keplerian disk (magnetic slab) in which the material is confined by the magnetic field of the accretion flow itself. This scenario, which we refer to as the Magneto-Levitation Accretion (MLA), has previously been developed for the case of accretion onto a black hole (see, e.g., Refs.\,\refcite{Shvartsman-1971}--\refcite{Igumenshchev-etal-2003}). An application of MLA scenario to the case of accretion onto a neutron star (see Refs.\,\refcite{Ikhsanov-Beskrovnaya-2012}--\refcite{Ikhsanov-etal-2013}) has led us to a number of important conclusions. First, the observed high X-ray luminosity of pulsars can be explained provided the mode by which the accreting material enters the stellar field from the magnetic slab is the anomalous (Bohm) diffusion. This indicates that plasma entry into the magnetosphere of both a neutron star and  the Earth and other planets is governed by a similar mechanism. Accretion onto the stellar surface at a required rate within the MLA scenario would occur even if the interchange instabilities of the magnetospheric boundary are suppressed. Second, the magnetospheric radius of the neutron star within the MLA scenario is a factor of a few smaller than the magnetospheric radius evaluated within the traditional quasi-spherical or Keplerian disk scenarios. Finally, the maximum possible spin-down torque applied to a neutron star from the magnetic slab is an order of magnitude larger that the upper limit to the spin-down torque expected in the traditional accretion scenarios and can reach the value of the spin-down torque inferred from observations. These basic conclusions are illustrated in the following Sections.

 \section{Spin-down rates}\label{2}

Parameters of five best studied long-period X-ray pulsars (with the period, $P_{\rm s}$, in excess of 100\,s) are listed in Table\,\ref{t1}. All of them are identified with HMXBs with the orbital period $P_{\rm orb}$. The X-ray luminosity of the pulsars, $L_{36}$, is given in units $10^{36}\,{\rm erg\,s^{-1}}$. The surface magnetic field of the neutron star, $B_{12} = B_*/10^{12}$\,G, is measured through observations of the cyclotron line in its spectrum. The last two columns show the spectral type of the massive component and the source distance.

\begin{table}[ph]
\tbl{Selected Long-period X-ray Pulsars (LPXPs)}
{\begin{tabular}{lccccccc}
\toprule
 Name ~&~
 $P_{\rm s}$,\,s &
 $P_{\rm orb}$,\,d &
 $L_{36},\,{\rm erg\,s^{-1}}$ &
 $B_{12}$,\,G &
 Normal &
 d,\,kpc &
 Ref. \\ \colrule
%
% \noalign{\smallskip}
% \hline
% \noalign{\smallskip}
%
%
 Vela~X--1 &
 283  &
 9    &
 4    &
 2.6  &
 B0.5~Ib &
 2    &
 \refcite{Nagase-etal-1986},\refcite{Kreykenbohm-etal-2002} \\
 4U~1907+09  &
 441   &
 8     &
 2     &
 2.1   &
 O8-9~Ia &
 4     &
 \refcite{Cox-etal-2005},\refcite{Cusumano-1998} \\
 4U~1538--52 &
 525   &
 4     &
 2     &
 2.3   &
 B0~Iab &
 4.5   &
 \refcite{Parkes-etal-1978},\refcite{Coburn-etal-2002} \\
 GX~301--2 &
 685   &
 41.5  &
 10    &
 4     &
 B1~Ia& 3  &
 \refcite{Chichkov-etal-1995},\refcite{Kreykenbohm-2004} \\
 X~Persei &
 837   &
 250   &
 0.1   &
 3.3   &
 B0~Ve& 1 &
\refcite{Palombara-2007},\refcite{Coburn-etal-2001} \\
 \botrule
\end{tabular}
 \label{t1}}
\end{table}

\begin{figure}[pb]
\centerline{\psfig{file=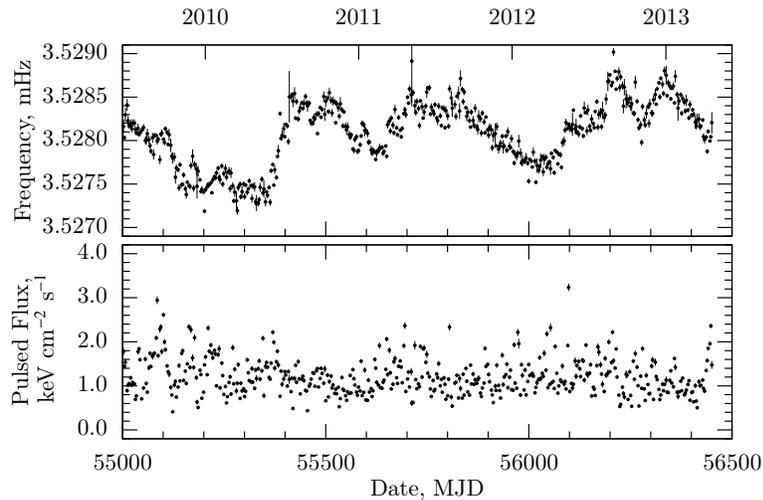,width=10cm}}
\vspace*{8pt}
\caption{Spin evolution (top) and $12-50$\,keV flux evolution (bottom) of Vela\,X-1
(Fermi/GMB observations:\,http://gammaray.msfc.nasa.gov/gbm/science/pulsars/lightcurves/)
\label{f1}}
 \end{figure}

An example of typical spin evolution of LPXPs is shown in Fig.\,\ref{f1}. The spin-up events alternate spin-down trends while the average period  does not change significantly on the 10--20\,yr timescale. The highest spin-down rates of the selected pulsars are listed in Table\,\ref{t2}. It shows Julian dates ($t_0$ and $t_1$), the total duration, $\bigtriangleup t$, and the absolute value of the spin-down rate, $|\dot{\nu}_{\rm sd}^{\rm obs}|$, inferred from observations of  spin-down events.

\begin{table}[ht]
\tbl{Observed spin-down rates of selected LPXPs}
{\begin{tabular}{lccccc}
\toprule
 Name &
 $t_0$,\,MJD &
 $t_1$,\,MJD &
 $\bigtriangleup t$,\,d &
 $|\dot{\nu}_{\rm sd}^{\rm obs}|,\,{\rm Hz\,s^{-1}}$ &
 Ref. \\ \colrule
Vela~X--1  & 44306 & 44320 &   14& $3\times10^{-13}$ & \refcite{Deeter-etal-1987} \\
4U~1907+09    & 54280 & 55600 & 1320& $4\times10^{-14}$ & \refcite{Sahiner-etal-2012} \\
4U~1538--52 & 45514 & 45522 &    8& $2\times10^{-13}$ & \refcite{Makishima-etal-1987}\\
GX~301--2     & 54300 & 54710 &  410 & $10^{-13}$ & \refcite{Evangelista-etal-2010}\\
X~Persei     & 43413 & 43532 &  118& $2\times10^{-14}$ & \refcite{Weisskopf-etal-1984} \\
  \botrule
\end{tabular}
 \label{t2}}
 \end{table}

\section{Spin-down torque}

The equation governing spin evolution of an accreting neutron star reads
 \be\label{main}
 2 \pi I \dot{\nu} = K_{\rm su} - K_{\rm sd},
 \ee
where $I$ is the moment of inertia, $\dot{\nu} = d\nu/dt$ and $\nu = 1/P_{\rm s}$ is the rotation frequency of the neutron star. $K_{\rm su}$ and $K_{\rm sd}$ are the spin-up and spin-down torques applied to the star from the accreting material. As follows from this equation the spin-down rate of the star is limited to $|\dot{\nu}_{\rm sd}| \leq |K_{\rm sd}|/(2 \pi I)$.

The spin-down torque applied to a neutron star from the accretion flow in the general case can be evaluated as (see Refs.\,\refcite{Ikhsanov-Beskrovnaya-2012} -- \refcite{Ikhsanov-Beskrovnaya-2013})
 \be\label{ksd}
 |K_{\rm sd}(r_{\rm m})| = k_{\rm t}\ \frac{\mu^2}{\left(r_{\rm m} r_{\rm cor}\right)^{3/2}},
 \ee
where $k_{\rm t}$ is a dimensionless parameter of an order of unity. The dipole magnetic moment, $\mu = (1/2) B_*\,R_{\rm ns}^3$, and the corotation radius, $r_{\rm cor} = \left(GM_{\rm ns}/\omega_{\rm s}^2\right)^{1/3}$, of the neutron star can be derived from observations (see Table\,\ref{t1}). Here $M_{\rm ns}$ is the mass, $\omega_{\rm s} = 2 \pi/P_{\rm s}$ is the angular velocity and $R_{\rm ns}$ is the radius of the neutron star. The magnetospheric radius, $r_{\rm m}$, is a free parameter. Its value depends on the structure and physical conditions in the accretion flow. If the star undergoes spherical or Keplerian disk accretion at the rate $\dmf$ the radius of its magnetosphere is close to the Alfv\'en radius \cite{Arons-Lea-1976,Arons-1993} $r_{\rm A} = \left(\mu^2/\dmf\,(GM_{\rm ns})^{1/2}\right)^{2/7}$, which is defined by equating the ram pressure of the free-falling gas with the magnetic pressure due to dipole field of the neutron star. Putting $r_{\rm m} \geq r_{\rm A}$ to Eq.~(\ref{ksd}) one finds $|K_{\rm sd}(r_{\rm A})| = |K_{\rm sd}^{(0)}| \leq k_{\rm t}\dmf\,\omega_{\rm s}\,r_{\rm A}^2$. This result is in  agreement with the value of spin-down torque reported in previous investigations (see, e.g., Ref.\,\refcite{Shakura-etal-2012} and references therein) and indicates that the maximum possible spin-down rate of a neutron star accreting material from a quasi-spherical flow or Keplerian disk is limited to $|\dot{\nu}_{\rm sd}^{\rm (0)}| = |K_{\rm sd}^{\rm (0)}|/(2 \pi I)$. The value of the ratio $\dot{\nu}_{\rm sd}^{\rm (0)}/\dot{\nu}_{\rm sd}^{\rm obs}$ for the parameters of the selected pulsars is given in the second column of Table\,\ref{t3}. It shows that the observed spin-down rate of these pulsars is an order of magnitude higher than the maximum possible value of the spin-down rate predicted within the traditional accretion scenarios.

Solving inequality $|\dot{\nu}_{\rm sd}^{\rm obs}| \leq |K_{\rm sd}(r_{\rm m})|/(2 \pi I)$ for $r_{\rm m}$ one finds that the pulsars would brake at the observed rate if the magnetospheric radius satisfies the condition $r_{\rm m} \leq r_0$, where
 \be\label{r0}
 r_0 = \frac{1}{r_{\rm cor}} \left(\frac{k_{\rm t} \mu^2} {2 \pi I \dot{\nu}_{\rm sd}}\right)^{2/3}.
 \ee
The values of $r_0$ and $r_{\rm A}$ as well as the ratio $r_0/r_{\rm A}$ for the parameters of the selected pulsars are given in the last three columns of Table\,\ref{t3}. It shows that the observed spin evolution of these pulsars can be explained provided the magnetospheric radius of the neutron star is substantially smaller than the canonical Alfv\'en radius. This situation is realized in the MLA scenario.

\begin{table}[ph]
\tbl{Parameters of the neutron star in LPXPs: the ratio of the observed ($\dot{\nu}_{\rm sd}^{\rm obs}$) to predicted spin-down rate in quasi-spherical ($\dot{\nu}_{\rm sd}^{\rm (0)}$) and MLA ($\dot{\nu}_{\rm sd}^{\rm (sl)}$) scenarios; and magnetospheric radius in comparison with value inferred from observations (see text for details)}
{\begin{tabular}{lccccc}
\toprule
 Name~&$\dot{\nu}_{\rm sd}^{\rm (0)}/\dot{\nu}_{\rm sd}^{\rm obs}$ ~&~
$\dot{\nu}_{\rm sd}^{\rm (sl)}/\dot{\nu}_{\rm sd}^{\rm obs}$
 &~ $r_0$,\,cm &~
 $r_A$,\,cm~ &~
 $r_0/r_A$ \\ \colrule
Vela~X--1  & 0.08 & 2.7  & $1.3\times10^{8}$ & $5.6\times10^{8}$ & 0.24\\
4U~1907+09   &  0.24 & 7.2  & $2.9\times10^{8}$ & $6.0\times10^{8}$ & 0.48\\
4U~1538--52 &  0.04 & 1.3  & $9.5\times10^{7}$ & $6.3\times10^{8}$ & 0.15\\
GX~301--2 &  0.24 & 8.4    & $2.6\times10^{8}$ & $5.5\times10^{8}$ & 0.48\\
X~Persei  &  0.09 & 2.7    & $4.6\times10^{8}$ & $1.8\times10^{9}$ & 0.25\\
  \botrule
\end{tabular}
 \label{t3}}
 \end{table}

 \section{Magneto-Levitation Accretion (MLA)}

The MLA scenario in a HMXB can be realized if the material which the neutron star captures from the wind of its massive companion is magnetized. The initial quasi-spherical flow in this case decelerates by its own magnetic field at a so called Shvartsman radius\cite{Shvartsman-1971},
 \be
R_{\rm sh} = \beta_0^{-2/3} \left(\frac{c_{\rm s}(r_{\rm G})}
 {v_{\rm rel}}\right)^{4/3} r_{\rm G},
 \ee
and is converted into a slowly rotating disk (magnetic slab) in which the material is confined by the magnetic field of the accretion flow itself (see Refs.\,\refcite{Bisnovatyi-Kogan-Ruzmaikin-1974} -- \refcite{Igumenshchev-etal-2003}). Here $\beta_0$ is the ratio of the thermal, $\rho c_{\rm s}^2$, to magnetic, $B_{\rm f}^2/8\pi$, energy of the accretion flow at the Bondi radius, $r_{\rm G} = 2GM_{\rm ns}/v_{\rm rel}^2$, of the neutron star which moves through the stellar wind with the relative velocity $v_{\rm rel}$, and $B_{\rm f}$ is the intrinsic magnetic field of the accretion flow.

The magnetospheric radius of a neutron star accreting material from the magnetic slab can be defined by the following basic conditions (see Refs.\,\refcite{Ikhsanov-2012},\refcite{Ikhsanov-etal-2013}):
 \begin{itemlist}
 \item the magnetic pressure due to dipole field of the neutron star at the magnetospheric boundary is equal to the external gas pressure, i.e. $\displaystyle\frac{\mathstrut \mu^2}{2 \pi r_{\rm m}^6} = \rho(r_{\rm m}) c_{\rm s}^2(r_{\rm m})$, and
 \item the rate of plasma diffusion into the field at the magnetospheric boundary is equal to the mass accretion rate onto the neutron star, i.e. $\dmf_{\rm in}(r_{\rm m}) = \displaystyle\frac{\mathstrut L_{\rm X} R_{\rm ns}}{GM_{\rm ns}}$.
 \end{itemlist}
Assuming the plasma entry into the field is governed by the anomalous (Bohm) diffusion one finds the magnetospheric radius of the neutron star in the form (for discussion see, e.g., Refs.\,\refcite{Ikhsanov-Beskrovnaya-2012} -- \refcite{Ikhsanov-etal-2013})

 \be\label{rma}
 r_{\rm ma} = \left(\frac{c\,m_{\rm p}^2}{16\,\sqrt{2}\,e\,k_{\rm B}}\right)^{2/13} \frac{\mu^{6/13} (GM_{\rm ns})^{5/13}}{T_0^{2/13} L_{\rm X}^{4/13} R_{\rm ns}^{4/13}}.
 \ee
Here $m_{\rm p}$ is the proton mass, $e$ is the electric charge of an electron, $k_{\rm B}$ is the Boltzmann constant and $T_0$ is the gas temperature in the region of interaction between the slab and the stellar field (magnetopause).

The maximum possible spin-down rate of a neutron star expected within the MLA scenario can be evaluated as
 \be
 |\dot{\nu}_{\rm sd}^{\rm (sl)}| \leq \frac{|K_{\rm sd}(r_{\rm ma})|}{2 \pi I}.
 \ee
The ratio $\dot{\nu}_{\rm sd}^{\rm (sl)}/\dot{\nu}_{\rm sd}^{\rm obs}$ for the parameters of the selected pulsars is listed in the third column of Table\,\ref{t3}. Thus, the observed spin evolution of the selected long-period pulsars can be explained within the MLA scenario provided $k_{\rm t} \sim 0.1-0.76$.

\section*{Acknowledgments}

NRI thanks MPIK at Heidelberg for kind hospitality. This investigation has been partly supported by the Alexander von Humboldt Foundation, Russian Foundation of Basic Research under the grant Nr.\,13-02-00077 and the Program of the Presidium of RAS Nr.\,21.

\end{document}